\documentclass[thmsa,a4paper,oneside,onecolumn,12pt,final,notitlepage]{article}
\usepackage{amssymb}

\usepackage{sw20lart}

\input{tcilatex}
\begin{document}

\title{Common Cause and Contextual Realization of Bell Correlation}
\author{A. Shafiee\thanks{%
E-mail: shafiee@theory.ipm.ac.ir} $^{\text{(1)}}$, R. Maleeh $^{\text{(2)}}$%
\ and \ M. Golshani $^{\text{(3,4)}}\bigskip $\quad \\
{\small \ }$\stackrel{1)}{}$ {\small Department of Chemistry, Sharif
University of Technology,}\\
{\small \ P.O.Box 11365-9516, Tehran, Iran}\\
$\stackrel{2)}{}$ {\small Department of Philosophy of Science, Sharif
University of Technology,}\\
{\small \ P.O.Box 11365-9161, Tehran, Iran.}\\
{\small \ }$\stackrel{3)}{}$ {\small Department of Physics, Sharif
University of Technology,}\\
{\small \ P.O.Box 11365-9161, Tehran, Iran.}\\
{\small \ }$\stackrel{4)}{}$ {\small Institutes for Studies in Theoretical
Physics \& Mathematics,}\\
{\small \ P.O.Box 19395-5531, Tehran, Iran.}}
\maketitle

\begin{abstract}
Considering the common cause principle, we construct a local-contextual
hidden-variable model for the Bohm version of EPR experiment. Our proposed
model can reproduce the predictions of quantum mechanics. It can be also
extended to classical examples in which similar correlations may be revealed.
\end{abstract}

\section{Introduction}

According to Bell's Theorem [1], one cannot construct a local realistic
hidden-variable theory that can reproduce all the statistical predictions of
quantum mechanics for a two-particle singlet-state. Here, by the term
locality we mean Bell's locality condition which in a local stochastic
hidden-variable theory is equivalent to the statistical independence of the
values of the spin components of the two particles [2]. Bell's locality
condition at the level of hidden variables also implies a more general tenet
named the ``common cause principle'' [3]. According to this principle, one
can explain how two spatially separated particles can be correlated, when
there is no direct causal relationship between them. Despite the fact that
one cannot decisively close the door to other potential alternatives which
may provide a better understanding of the so-called Bell-type correlations
in a causal fashion [4], the common cause description is still a very
attractive and comprehensible way of understanding the nature of these
correlations [5]. But, thanks to Bell's theorem, we do know that this
description cannot be compatible with the predictions of quantum mechanics.
So, we are still facing the crucial problem of whether correlations
resulting from the entanglement of two spatially separated particles are an
essential characteristic of microphysical reality. What is then the nature
of these correlations? Could they still be explained by invoking a field of
common cause in the past light cones of the two separated particles?

The main purpose of this article is to give a positive answer to the last
question. Here, we want to provide a stochastic hidden-variable model which
honors the common cause principle but cannot be ruled out by Bell's theorem.
The key point here is that it is a local \textit{contextual }model. By the
term contextual we mean that each probability function is defined in a
measurement setup where a definite set of hidden variables is used. This
approach is very similar to what has been recently suggested by de Muynck
[6]. Here, we are going to provide a more concrete realization of what de
Muynck has proposed as contextual reality which might be accomplished by
some hidden variables in a subquantum theory.

In section 2, we first review the common cause principle in the domain of
probabilistic causality. Then, in section 3, we construct a local contextual
hidden-variable model which confirms the quantum mechanical predictions. In
section 4, we explain why our proposed model is local and contextual.
Subsequently, we talk about an alternative classical example in section 5
which can make the subtleties of our argument more transparent. We have
summarized our results in the last section.

\section{\textbf{Common Cause Principle}}

It seems better to explain common cause principle through the term
''spurious correlation''. Eells defines the usual understanding of this term
in the following way [7]:

\begin{quote}
''\textit{[T]wo factors are spuriously correlated when (roughly) neither
causes the other and the correlation disappears when a third variable is
introduced and ''held fixed''- that is, the correlation disappears both in
the presence and in the absence of the third factor}.''
\end{quote}

If a correlation between factors $x$ and $y$ disappears both in the presence
and in the absence of third factor $z$, then the explanation may be either
that the correlation results from the joint causal effect of $z$ on $x$ and $%
y$ ($z$ is a common cause of $x$ and $y$) or that $z$ is an intermediate
causal factor between $x$ and $y$ ($x$ operates on $y$ through $z$ or $y$
operates on $x$ through $z$). Since one of $x$ and $y$ may in fact be a
genuine positive causal factor for the other in the second case, we do not
consider it as a case of spurious correlation.

So, as it is indicated by Eells, two factors $x$ and $y$ are spuriously
correlated if neither causes the other and the correlation of $y$ with $x$
disappears when $z$ is held fixed. Now, suppose that, in the simplest kind
of a common cause pattern, $z$ is a common cause of both $x$ and $y.$ Then,
the following relations hold for the various kinds of probabilities:

\begin{equation}
\Pr (x|z)>\Pr (x|\thicksim z)  \tag{1}
\end{equation}

\begin{equation}
\Pr (y|z)>\Pr (y|\thicksim z)  \tag{2}
\end{equation}

\begin{equation}
\Pr (y|z\&x)=\Pr (y|z\&\thicksim x)  \tag{3}
\end{equation}

\begin{equation}
\Pr (y|\thicksim z\&x)=\Pr (y|\thicksim z\&\thicksim x)  \tag{4}
\end{equation}

\begin{equation}
\Pr (x\&y|z)=\Pr (x|z)\Pr (y|z)  \tag{5}
\end{equation}

\begin{equation}
\Pr (x\&y|\thicksim z)=\Pr (x|\thicksim z)\Pr (y|\thicksim z)  \tag{6}
\end{equation}
where $\thicksim $ denotes the negation. Relations (1) and (2) correspond to
the assumption that $z$ is a common cause of $x$ and $y$; (3) and (4) say
that the correlation between $x$ and $y$ disappears when $z$ is held fixed
(positively or negatively) and (5) as well as (6) mean that when $z$ is held
fixed (positively or negatively), the joint probability of $x$ and $y$ will
be factorized.

\section{\textbf{A Local Contextual Hidden-Variable Model }}

Let us consider an ideal version of EPR-Bohm experiment in which a spin zero
source emits two entangled spin $\frac{1}{2}$ particles in the singlet-state
[8]. Alice (Bob) measures the spin component of particle $1$ ($2$) along the
direction $\widehat{a}$ ($\widehat{b}$), denoted by $S_{a}^{(1)}$ $%
(S_{b}^{(2)})$ which takes the values $r=\pm 1$ $(q=\pm 1)$ in units of $%
\dfrac{\hslash }{2}$. The directions $\widehat{a}$ and $\widehat{b}$ lie on
the $y$-$z$ plane. Now, we consider a hidden-variable model in which we have
complete information about the spin component of each particle along any
direction, given that the corresponding hidden variables along the same
direction are known. Moreover, this model enables us to make some
statistical predictions along the other directions. The framework of our
proposed hidden-variable model consists of the following suppositions:

\begin{quotation}
\textbf{S1}. The set of hidden variables $\lambda _{j,-j}^{n}$ ($\widehat{n}=%
\widehat{a},\widehat{b},....$; $j=\pm 1$ in units of $\dfrac{\hslash }{2}$)
constitute the field of common cause, where $\widehat{n}=\sin \theta $ $%
\widehat{y}+\cos \theta $ $\widehat{z}$ is a unit vector in the $y$-$z$
plane. Taking into account the conservation principle of spin angular
momentum, $\lambda _{j,-j}^{n}$ uniquely determine the values $j$ and $-j$
for the spin components of particles $1$ and $2$, respectively, along the
direction $\widehat{n}$. The outcomes of any spin measurement along $%
\widehat{n}$ are also assumed to be \textit{predetermined} by $\lambda
_{j,-j}^{n}$.

\textbf{S2}. The spin component $S_{\lambda _{_{j,-j}}^{n}}^{(i)}=\pm 1$ is
assumed to contain complete information about the spin value of particle $i$
($i=1$ or $2$) along the direction $\widehat{n}.$ Subsequently, we define a
spin vector along $\widehat{n}$ which is a unit vector, i.e.,

\begin{equation}
\overrightarrow{S_{\lambda _{_{j,-j}}^{n}}^{(1)}}=j\widehat{n}\text{ \ and\ }%
\overrightarrow{S_{\lambda _{j,-j}^{n}}^{(2)}}=-j\widehat{n}  \tag{7}
\end{equation}

\textbf{S3}. There is an equal chance for the hidden variables $\lambda
_{+-}^{n}$ and $\lambda _{-+}^{n}$ to occur in a given run of the
experiment. So, we are assuming that:

\begin{equation}
\Pr (\lambda _{+-}^{n})=\Pr (\lambda _{-+}^{n})=\frac{1}{2}  \tag{8}
\end{equation}
where $\Pr $ stands for probability.

\textbf{S4}. Regarding the stochastic character of the model, if $\lambda
_{j,-j}^{n}$ are given, we shall have only information about the \textit{mean%
} value of the spin component of particle $i$ along another direction $%
\widehat{n^{\prime }}\neq \widehat{n}$, which is denoted by $E^{(i)}(\lambda
_{j,-j}^{n},\widehat{n^{\prime }})$. The mean value is postulated to be
obtained by projecting the spin vector $\overrightarrow{S_{\lambda
_{_{j,-j}}^{n}}^{(i)}}$ on the purposed direction. That is,

\begin{equation}
E^{(i)}(\lambda _{j,-j}^{n},\widehat{n^{\prime }})=\overrightarrow{%
S_{\lambda _{_{j,-j}}^{n}}^{(i)}}.\widehat{n^{\prime }}  \tag{9}
\end{equation}
\end{quotation}

Applying \textbf{S4} for $\widehat{n}=\widehat{n^{\prime }}$, one gets

\begin{equation}
E^{(1)}(\lambda _{j,-j}^{n},\widehat{n})=j,\text{ }E^{(2)}(\lambda
_{j,-j}^{n},\widehat{n})=-j;  \tag{10}
\end{equation}
which means that we have complete information along the direction $\widehat{n%
}$, given that $\lambda _{j,-j}^{n}$ is known. Furthermore, Using \textbf{S2 
}and \textbf{S4}, one can also show that, e.g.,

\begin{eqnarray}
E^{(2)}(\lambda _{j,-j}^{n},\widehat{n^{\prime }}) &=&\overrightarrow{%
S_{\lambda _{_{j,-j}}^{n}}^{(2)}}.\widehat{n^{\prime }}=-j\widehat{n}.%
\widehat{n^{\prime }}  \nonumber \\
&=&-j\cos \varphi _{nn^{\prime }}  \tag{11}
\end{eqnarray}
where $\varphi _{nn^{\prime }}=\mid \widehat{n}-\widehat{n^{\prime }}\mid $.
Similarly, $E^{(1)}(\lambda _{j,-j}^{n},\widehat{n^{\prime }})=j\cos \varphi
_{nn^{\prime }}$.

Considering \textbf{S1 }and \textbf{S2}, when Alice measures the spin
component of the first particle along $\widehat{a}$ and obtains a result $r$%
, it will be obvious for her that the hidden variables $\lambda _{r,-r}^{a}$
have caused such a result. She can also infer that if Bob measures along the
same direction, he will obtain the result $-r$. However, Alice cannot
predict with certainty that if Bob measures along a different direction $%
\widehat{b}\neq \widehat{a}$, what result he will obtain. Nevertheless,
using the assumptions \textbf{S1}, \textbf{S2} and \textbf{S4}, after her
measurement, Alice can refer to $\lambda _{r,-r}^{a}$ to have some
statistical predictions at a subquantum level. For example, according to the
relation (11) she can predict that the mean value of Bob's results will be $%
E^{(2)}(\lambda _{r,-r}^{a},\widehat{b})=-r\cos \varphi _{ab}$, if he
measures along $\widehat{b}$\footnote{%
Indeed, this is only the analytical form of the prediction, not its
numerical one. To have knowledge of the latter case, one needs an exact
knowledge of $\varphi _{ab}$ too.}. Albeit this result is not observable,
one should average over the hidden variables $\lambda _{r,-r}^{a}$ to obtain
the empirical results (see below).

The same situation holds for Bob: After his measurement along $\widehat{b}$
and getting the result $q$, he concludes that the mean value of Alice's
results will be definitely $E^{(1)}(\lambda _{-q,q}^{b},\widehat{a})=-q\cos
\varphi _{ab}$. (For $\widehat{a}=\widehat{b}$, $r=-q$, according to \textbf{%
S1}.) Since the choice of measuring direction can be arbitrarily made by
each local observer, different sets of hidden variables may be used by Alice
and Bob to make predictions about the results of the other one. These
predictions might be different at a subquantum level. The hidden variables,
however, constitute the same field of common cause. So, each observer can
refer to a definite set of hidden variables to reproduce the \textit{%
observational} predictions. Their predictions should be the same,
experimentally. Hence, to reproduce the Bell correlations, Alice can use the
local hidden variables $\lambda _{r,-r}^{a}$, when she reveals the result $r$
for $S_{a}^{(1)}$, which we call hereafter the Alice description. Bob can
also invoke $\lambda _{-q,q}^{b}$, when he reveals the result $q$ for $%
S_{b}^{(2)}$, which we call the Bob description (Fig 1.) As we shall show in
the following, the observational results in both descriptions are the same
and they are supposed to be equal to the predictions of quantum mechanics.
In the following, we first use the Alice description.

To see some of the predictions of this model at the level of hidden
variables, it is apparent from \textbf{S1 }and \textbf{S2 }that:

\begin{equation}
\Pr (S_{a}^{(1)}=r,S_{a}^{(2)}=-r\mid \lambda _{r,-r}^{a},\widehat{a},%
\widehat{a})=1;\ \Pr (S_{a}^{(1)}=r,S_{a}^{(2)}=r\mid \lambda _{r,-r}^{a},%
\widehat{a},\widehat{a})=0  \tag{12}
\end{equation}
Then, due to the factorizability of the joint probabilities in (5), in the
presence of the common cause (supposing in \textbf{S1}), one can show that

\begin{equation}
\Pr (S_{a}^{(1)}=r,S_{b}^{(2)}=q\mid \lambda _{r,-r}^{a},\widehat{a},%
\widehat{b})=\Pr (S_{a}^{(1)}=r\mid \lambda _{r,-r}^{a},\widehat{a})\Pr
(S_{b}^{(2)}=q\mid \lambda _{r,-r}^{a},\widehat{b})  \tag{13}
\end{equation}
Here, $\Pr (S_{a}^{(1)}=r\mid \lambda _{r,-r}^{a},\widehat{a})=1$ and

\begin{equation}
\Pr (S_{b}^{(2)}=q\mid \lambda _{r,-r}^{a},\widehat{b})=\frac{1}{2}\left[
1+qE^{(2)}(\lambda _{r,-r}^{a},\widehat{b})\right] =\frac{1}{2}\left[
1-rq\cos \varphi _{ab}\right]  \tag{14}
\end{equation}
where the last equality is obtained by using the relation (11).

We can also calculate the correlation function both at the level of
hidden-variables and quantum mechanics. The correlation function at the
level of hidden variables is equal to:

\begin{equation}
C_{\lambda _{r,-r}^{a}}(S_{a}^{(1)},S_{b}^{(2)})=E^{(12)}(\lambda
_{r,-r}^{a},\widehat{a},\widehat{b})-E^{(1)}(\lambda _{r,-r}^{a},\widehat{a}%
)E^{(2)}(\lambda _{r,-r}^{a},\widehat{b})  \tag{15}
\end{equation}
where $E^{(12)}(\lambda _{r,-r}^{a},\widehat{a},\widehat{b})$ is the average
value of the spin components of the two particles along $\widehat{a}$ and $%
\widehat{b}$, and it is defined as:

\begin{equation}
E^{(12)}(\lambda _{r,-r}^{a},\widehat{a},\widehat{b})=\stackunder{k,l=\pm 1}{%
\sum }kl\Pr (S_{a}^{(1)}=k,S_{b}^{(2)}=l\mid \lambda _{r,-r}^{a},\widehat{a},%
\widehat{b})  \tag{16}
\end{equation}
Using the relations (13) and (14) in (16), one gets:

\begin{equation}
E^{(12)}(\lambda _{+-}^{a},\widehat{a},\widehat{b})=E^{(12)}(\lambda
_{-+}^{a},\widehat{a},\widehat{b})=-\cos \varphi _{ab}  \tag{17}
\end{equation}

Accordingly, we reach the following result:

\begin{equation}
C_{\lambda _{r,-r}^{a}}(S_{a}^{(1)},S_{b}^{(2)})=-\cos \varphi
_{ab}-r(-r\cos \varphi _{ab})=0  \tag{18}
\end{equation}
which means that there is no correlation at the subquantum level, as was
expected.

In quantum mechanics, the correlation function is equal to:

\begin{equation}
C_{QM}(S_{a}^{(1)},S_{b}^{(2)})=\left\langle
S_{a}^{(1)}S_{b}^{(2)}\right\rangle -\left\langle S_{a}^{(1)}\right\rangle
\left\langle S_{b}^{(2)}\right\rangle  \tag{19}
\end{equation}
where $\left\langle ...\right\rangle $ stands for the quantum mechanical
expectation values. Using \textbf{S3}, one can show that

\begin{eqnarray}
\left\langle S_{b}^{(2)}\right\rangle &=&\Pr (\lambda
_{+-}^{a})E^{(2)}(\lambda _{+-}^{a},\widehat{b})+\Pr (\lambda
_{-+}^{a})E^{(2)}(\lambda _{-+}^{a},\widehat{b})  \nonumber \\
&=&\frac{1}{2}(-r\cos \varphi _{ab})+\frac{1}{2}(+r\cos \varphi _{ab})=0 
\tag{20}
\end{eqnarray}
where we have used the relation (11) for $\widehat{n}=\widehat{a}$ and $%
\widehat{n^{\prime }}=\widehat{b}$. Similarly one can show that $%
\left\langle S_{a}^{(1)}\right\rangle =0$. We also have

\begin{eqnarray}
\left\langle S_{a}^{(1)}S_{b}^{(2)}\right\rangle &=&\Pr (\lambda
_{+-}^{a})E^{(12)}(\lambda _{+-}^{a},\widehat{a},\widehat{b})+\Pr (\lambda
_{-+}^{a})E^{(12)}(\lambda _{-+}^{a},\widehat{a},\widehat{b})  \nonumber \\
&=&\frac{1}{2}(-\cos \varphi _{ab})+\frac{1}{2}(-\cos \varphi _{ab})=-\cos
\varphi _{ab}  \tag{21}
\end{eqnarray}
in which the relation (17) is used. Relations (20) and (21) lead us to the
following result:

\begin{equation}
C_{QM}(S_{a}^{(1)},S_{b}^{(2)})=-\cos \varphi _{ab}  \tag{22}
\end{equation}
which is the correlation function predicted by quantum mechanics\textbf{.}

The same results could be obtained, if we used the Bob description. For
example, one could replace the left hand side of relation (14) with $\Pr
(S_{a}^{(1)}=r\mid \lambda _{-q,q}^{b},\widehat{a})$ and get the same result
as $\frac{1}{2}\left[ 1-rq\cos \varphi _{ab}\right] .$ Here, $\Pr
(S_{b}^{(2)}=q\mid \lambda _{-q,q}^{b},\widehat{b})=1$ and the relations
(17) to (22) could be derived again, if one substitutes $\lambda _{-q,q}^{b}$
in place of $\lambda _{r,-r}^{a}.$ This shows that each local observer by
herself (himself) can simulate Bell correlations without any communication
with her (his) distant partner. So, there is no communication cost for
simulating Bell correlations in our approach [4]. This demonstrates the
correct application of a common cause description in an appropriate manner.

\section{Locality versus Contextuality}

Our proposed model is an instance of a local contextual hidden-variable
theory. Contextulaity in our definition means that we should always refer to
a definite set of hidden variables, in a given measuring setup, to be able
to have some observational predictions. Hence, the joint probability of $%
S_{a}^{(1)}$ and $S_{b}^{(2)}$ being $r$ and $q$ respectively, in a context
which is characterized by the hidden variables $\lambda _{r,-r}^{a}$ and the
measuring directions $\widehat{a}$ and $\widehat{b}$ in Alice description
(as was denoted by $\Pr (S_{a}^{(1)}=r,S_{b}^{(2)}=q\mid \lambda _{r,-r}^{a},%
\widehat{a},\widehat{b})$ in (13)), is essentially different from $\Pr
(S_{a}^{(1)}=r,S_{b}^{(2)}=q\mid \lambda _{-r,r}^{a},\widehat{a},\widehat{b}%
) $ and/or $\Pr (S_{a^{\prime }}^{(1)}=r,S_{b}^{(2)}=q\mid \lambda
_{r,-r}^{a^{\prime }},\widehat{a^{\prime }},\widehat{b})$ where in the
latter case, Alice measures the spin component of the first particle along a
different direction $\widehat{a^{\prime }}\neq \widehat{a}$. We are not
allowed to sum up conditional probabilities in different contexts where
distinctive hidden variables characterizes different conditionals.
Consequently, Bell's inequality cannot be derived because different sets of
hidden variables (including different pieces of information) should be used
in different contexts.

However, the model is obviously local. Alice and Bob refer to their own
local hidden variables $\lambda _{r,-r}^{a}$ and $\lambda _{-q,q}^{b}$,
respectively. There is no causal link between them. Each local observer
refers to her (his) context dependent hidden variables to make a proper
prediction. Notwithstanding, the predictions of distant observers are not
necessarily the same, whenever a hidden-variable description is taken into
account. They have no correlation at this level. For example, Alice can
attribute the probability $\Pr (S_{a}^{(1)}=r\mid \lambda _{r,-r}^{a},%
\widehat{a})=1$ to the outcome $S_{a}^{(1)}=r$, whereas Bob predicts $\Pr
(S_{a}^{(1)}=r\mid \lambda _{-q,q}^{b},\widehat{a})=\frac{1}{2}\left[
1-rq\cos \varphi _{ab}\right] $ for the same outcome at the same time.
Nevertheless, their observational predictions are always the same. The
spurious correlation appears when the local hidden variables are
statistically averaged.

It is also important to stress here that the statistical description of this
model is intrinsically realistic. While in quantum mechanics probabilities
do not describe the current situation of the system, here we can describe
the statistical behavior of the particles in a definite context, if the
corresponding hidden variables are known. For example, we can simultaneously
attribute probabilities to the spin components of particle $2$ along
different directions. That is, all spin components are real and the relation
(14) holds true for every $\widehat{b}$, given that $\lambda _{r,-r}^{a}$
are known at the same time. But in practice, since one cannot \textit{measure%
} $S_{b}^{(2)}$ for different directions simultaneously, we should use the
counterfactual statements for what could be described for different
directions in a given measuring setup. This means that the observational
predictions of this model for incompatible observables should be interpreted
counterfactually. This is different from the standard interpretation of
quantum mechanics in which the randomness of events is usually believed to
be irreducible. Instead, the quantum results are always reproduced in this
model, because we are \textit{ignorant} of which of the hidden variables,
e.g., $\lambda _{+-}^{a}$ or $\lambda _{-+}^{a}$ is applicable in different
runs of a given experiment.

\section{A Classical example}

We are now going to present a classical counterpart for a Bell-type
experiment which obeys similar rules as what was mentioned for a
local-contextual model in section 3. Let us assume that Alice and Bob are
two observers who are located at different positions with a space-like
separation. There is no communication between them. Intermediately, there is
a place where a sender (Sam) stores balls with different colors. On each
ball, one of the two signs $+$ or $-$ is inscribed. Sam sends balls with
different colors to each of the two observers in different stages of an
experiment successively. Each observer has a color-recognizer device which
is able to detect the passage of any colored ball but can be adjusted to
identify only \textit{one} type of color during each step of an experiment.
Choosing from Sam's colors set, let us assume here that Alice decides to
collect balls with color $a$ and Bob collects balls with color $b$, standing
for, e.g., amber and blue, respectively. They are not aware of which color
their partner chooses, nor Sam needs having knowledge of their decision.
Subsequently, Alice and Bob record the signs on their balls. Then, they will
upload their outcomes to a remote computer which can calculate the
correlation of data.

As a matter of fact, this experiment is composed of different stages.
Without loss of generality, one can suppose that Sam owns a set of three
colored balls including amber ($a$), blue ($b$) and cherry ($c$). In three
different stages, Sam sends both members of the pairs ($a,b$), ($a,c$) and ($%
c,b$) to Alice and Bob. For each pair of colored balls including ($a,b$), ($%
a,c$) and ($c,b$), Sam exploits specific executive algorithms which cause
definite correlations between the signs inscribed on balls. In each stage of
the experiment, Alice's (Bob's) device is capable of recognizing only one of
the two colors $a$ or $c$ ($b$ or $c$). The reader can find a full
description of our proposed experiment in the following Appendix.
Nevertheless, for our present purposes, let us consider a more restricted
situation where Alice and Bob choose $a$ and $b$ colors, respectively. So,
it is sufficient henceforth to focus on those events in which only $a$ and $%
b $ colored balls are detected simultaneously, and leave out the $c$ events.
In this way, Sam is committed to send these colors according to two
complementary executive algorithms which we call $\mathcal{A}_{1}$ and $%
\mathcal{A}_{2}$. Each algorithm includes some important features which are
described in the following:

\begin{quote}
1- Sam either sends $a$ colored balls with engraved symbols $+$ (hereafter
called as $a_{A}+$) or the same colored balls with engraved symbols $-$ ($%
a_{A}-$) along with $b$ colored balls (denoted by $b_{A}$) to Alice. The
first situation is controlled by $\mathcal{A}_{1}$ algorithm and the second
one by $\mathcal{A}_{2}$. Sam uses the two algorithms with equal chances,
i.e., $\Pr (\mathcal{A}_{1})=\Pr (\mathcal{A}_{2})=\frac{1}{2}$.

2- Suppose, for example, that Sam sends an $a_{A}+$ and a $b_{A}-$ ball to
Alice. Then, at the same time, he will surely send an $a_{B}-$ and a $b_{B}+$
ball to Bob. This means that Sam takes into account the anticorrelation rule
in choosing $\pm $ symbols for the same colored balls which are sent to
different observers. The anticorrelation rule is fulfilled by either of the $%
\mathcal{A}_{1}$ and $\mathcal{A}_{2}$ algorithms. Strictly speaking, e.g.,
under the $\mathcal{A}_{1}$ algorithm, $a_{A}+$ and $a_{B}-$ balls are
always sent to Alice and Bob, respectively. The $b$ colored balls should
also satisfy the anticorrelation rule at the same time.

3- When Sam executes the $\mathcal{A}_{1}$ algorithm, the probability that
Bob records $b_{B}+$ results is chosen to be 0.15. This means that Sam sends 
$b+$ balls to Bob (joined with $a+$ balls to Alice), with a frequency of 15
per 100 $b$ colored balls\footnote{%
Here, we are always assuming that the probability of an event is equivalent
to the frequency of occurrence of the same event. So, when we say that the
probability of a given outcome is, e.g., 0.15, we mean there are $N$ total
results for which our supposed outcome occures 0.15$N$ times.}. When Sam
uses the $\mathcal{A}_{2}$ algorithm, the probability that Bob records the
same outcome is 0.85 (i.e., a frequency of 85 per 100 $b$ colored balls).
Accordingly, when Alice records $a_{A}+$ ($a_{A}-$) results, Bob will
receive $b_{B}+$ ($b_{B}-$) balls with a probability of 0.15.

4- When Sam uses the $\mathcal{A}_{1}$ algorithm, each $b_{A}-$ ball is also
sent to Alice with a probability of 0.15. This is a consequence of the
anticorrelation rule in 2. Hence, according to $\mathcal{A}_{2}$ algorithm,
the probability that the same ball is sent to Alice is 0.85. This
accomplishes the stochastic characters of two executive algorithms used by
Sam.

5- This is an ideal experiment, in the sense that the color-recognizer
device is capable of registering any colored ball passing through it, even
if it cannot recognize its color. The experiment is ended when Alice and Bob
both have full registered data (i.e., the number of all color-registered
balls is compatible with all passing balls) for $a$ and $b$ colored balls,
respectively.
\end{quote}

Table 1 summarizes the executive routines of $\mathcal{A}_{1}$ and $\mathcal{%
A}_{2}$ in terms of some definite frequencies. Here, e.g., the results ($%
a_{A}+$, $b_{A}-$; $b_{B}+$, $a_{B}-$) denote that Sam sends simultaneously
a pair of $a$ colored balls with signs $+$ and $-$ as well as a pair of $b$
colored balls with opposite signs to Alice and Bob, respectively. (Here, we
are neglecting the $c$ colored balls, as was said before. For a full
description of this experiment see the following Appendix.) Under the $%
\mathcal{A}_{1}$, the probability (frequency) of occurrence of this event is
equal to:

\begin{eqnarray}
\Pr (a_{A}+,b_{A}-;b_{B}+,a_{B}-|\ \mathcal{A}_{1}) &=&\Pr
(a_{A}+;b_{B}+\mid \mathcal{A}_{1})\text{ }\Pr (b_{A}-;a_{B}-\mid \mathcal{A}%
_{1},a_{A}+,b_{B}+)  \nonumber \\
\ &=&\Pr (b_{A}-;a_{B}-\mid \mathcal{A}_{1})\text{ }\Pr (a_{A}+;b_{B}+\mid 
\mathcal{A}_{1},b_{A}-,a_{B}-)  \nonumber \\
&=&0.15\times 1=0.15  \tag{23}
\end{eqnarray}
The frequencies of other results can be defined similarly.\bigskip

\[
\begin{tabular}{|c|c|c|}
\hline
{\small Frequency}$\diagdown ${\small Results} & $%
(a_{A}+,b_{A}-;b_{B}+,a_{B}-)$ & $(a_{A}+,b_{A}+;b_{B}-,a_{B}-)$ \\ \hline
{\small Frequency according to }$\mathcal{A}_{1}$ & 0.15 & 0.85 \\ \hline
{\small Frequency}$\diagdown ${\small Results} & $%
(a_{A}-,b_{A}-;b_{B}+,a_{B}+)$ & $(a_{A}-,b_{A}+;b_{B}-,a_{B}+)$ \\ \hline
{\small Frequency according to }$\mathcal{A}_{2}$ & 0.85 & 0.15 \\ \hline
\end{tabular}
\]

\begin{center}
{\small Table 1. The executive routines of }$\mathcal{A}_{1}${\small \ and }$%
\mathcal{A}_{2}$ {\small in terms of some given probabilities.\bigskip }
\end{center}

The results indicated in table 1 are not exactly what Alice and Bob can
observe. As was said before, Alice only detects one of the two results $%
a_{A}+$ or $a_{A}-$. Correspondingly, Bob can only record $b_{B}+$ and $%
b_{B}-$ outcomes. So, what both observers can really observe is that they
are receiving their desired colors with an equal distribution of $+$ and $-$
signs. At the end of experiment, Alice and Bob upload their data to a remote
computer which is able to compare their results and calculate the
correlation of them. Table 2 demonstrates what the remote computer
calculates for the frequency of occurrence of the observed results and their
correlation.\bigskip

\[
\begin{tabular}{|c|cccc|}
\hline
{\small Statistical Calculations}$\diagdown ${\small Results} & $%
(a_{A}+;b_{B}+)$ & \multicolumn{1}{|c}{$(a_{A}+;b_{B}-)$} & 
\multicolumn{1}{|c}{$(a_{A}-;b_{B}+)$} & \multicolumn{1}{|c|}{$%
(a_{A}-;b_{B}-)$} \\ \hline
{\small Frequency } & $\frac{1}{2}\times 0.15$ & \multicolumn{1}{|c}{$\frac{1%
}{2}\times $ $0.85$} & \multicolumn{1}{|c}{$\frac{1}{2}\times 0.85$} & 
\multicolumn{1}{|c|}{$\frac{1}{2}\times $ $0.15$} \\ \hline
{\small Correlation} &  & $-0.7$ &  &  \\ \hline
\end{tabular}
\]

\begin{center}
{\small Table 2. The remote computer calculates the frequency of occurrence
of detected results and their correlation.\bigskip }
\end{center}

Here, for instance, the joint probability of getting an $a_{A}+$ result by
Alice in conjunction with a $b_{B}+$ result by Bob is equal to:

\begin{eqnarray}
\Pr (a_{A}+;b_{B}+) &=&\Pr (a_{A}+)\text{ }\Pr (b_{B}+\mid a_{A}+)  \nonumber
\\
\ &=&\Pr (b_{B}+)\text{ }\Pr (a_{A}+\mid b_{B}+)  \nonumber \\
&=&\frac{1}{2}\times 0.15  \tag{24}
\end{eqnarray}

Considering the outcomes of Alice and Bob collectively, the above result,
e.g., can be calculated directly by the remote computer. However, one can
also reach the same result, using Sam's information about the statistical
features of two executive algorithms $\mathcal{A}_{1}$ and $\mathcal{A}_{2}$
described above:

\begin{eqnarray}
\Pr (a_{A}+;b_{B}+) &=&\Pr (a_{A}+;b_{B}+\mid \mathcal{A}_{1})\Pr (\mathcal{A%
}_{1})+\Pr (a_{A}+;b_{B}+\mid \mathcal{A}_{2})\Pr (\mathcal{A}_{2}) 
\nonumber \\
&=&0.15\times \frac{1}{2}+0\times \frac{1}{2}=0.15\times \frac{1}{2} 
\tag{25}
\end{eqnarray}

Now let us see what Sam obtains for the correlations of the twofold results
in table 2 in the \textit{presence} of the determining factors $\mathcal{A}%
_{1}$ and $\mathcal{A}_{2}$ which should be viewed as two complementary
calculational contexts. Hence, one can define, e.g.:

\begin{equation}
C_{\mathcal{A}_{1}}(a_{A},b_{B})=\left\langle a_{A}\ b_{B}\right\rangle _{%
\mathcal{A}_{1}}-\left\langle a_{A}\right\rangle _{\mathcal{A}%
_{1}}\left\langle b_{B}\right\rangle _{\mathcal{A}_{1}}  \tag{26}
\end{equation}
where $C_{\mathcal{A}_{1}}(a_{A},b_{B})$ is the correlation function under
the $\mathcal{A}_{1}$ algorithm and the other functions are mean values
under the same program. From table 1, it is apparent that $\left\langle
a_{A}\ b_{B}\right\rangle _{\mathcal{A}_{1}}=-0.7$, $\left\langle
a_{A}\right\rangle _{\mathcal{A}_{1}}=+1$ and $\left\langle
b_{B}\right\rangle _{\mathcal{A}_{1}}=-0.7$. So, $C_{\mathcal{A}%
_{1}}(a_{A},b_{B})=0$. One can similarly show that $C_{\mathcal{A}%
_{2}}(a_{A},b_{B})$ is also equal to zero under the $\mathcal{A}_{2}$
algorithm. Consequently, \textit{the correlation disappears when a third
variable (}$\mathcal{A}_{1}$\textit{\ or }$\mathcal{A}_{2}$\textit{) is
introduced and held fixed. }Accordingly, the executive algorithms $\mathcal{A%
}_{1}$ and $\mathcal{A}_{2}$ constitute a field of common cause.

Comparing this with a Bell-type experiment, if one replaces the colored
balls $a$ and $b$ with the spin components of two particles along the
directions $\widehat{a}$ and $\widehat{b}$ respectively, it will be apparent
that the two executive algorithms $\mathcal{A}_{1}$ and $\mathcal{A}_{2}$ in
the present example stand for the hidden variables $\lambda _{+-}^{a}$ and $%
\lambda _{-+}^{a}$ in our Bell story. Here, however, we restricted ourselves
to a single measuring setup which includes only $a$ and $b$ colored balls.
So, $\mathcal{A}_{1}$ and $\mathcal{A}_{2}$ are solely defined in such a
measuring context. Yet, in such a context too, the two algorithms require
complementary calculational conditions under which one is not allowed to sum
up the probabilities directly. The expressions like $\Pr (...,...\mid 
\mathcal{A}_{1})+\Pr (...,...\mid \mathcal{A}_{2})$ are meaningless.
Alternatively, if one applies the probability rules correctly, one can hide
the role of different conditions. Correlation appears as an emergent result
of such a disguisement.

\section{Conclusion}

Bell's theorem states that there are no local hidden variables which can
screen off the mysterious correlations of two entangled particles. However,
it is still important to know what we can infer about the nature of these
correlations. Is there any non-local action? Or, do Alice and Bob only
update their state of knowledge, when they receive new information via the
classical channels of information? Here, the correlation appears when the
whole information is considered. A positive response to either of the above
questions introduces other problems which are regularly discussed in the
literature [9, 10, 11].

There is, however, another possibility when one introduces the notion of
contextuality. The consistency of the local contextual hidden-variable
theories is not necessarily ruled out by Bell's theorem, although this
really depends on what one means by the term contextual (for example, see
references [6, 12]). Here, we have tried to construct a local contextual
hidden-variable model which can reproduce the predictions of quantum
mechanics. The contextual character of this model requires that the type of
hidden variables used by two observers be held fixed according to their
corresponding measuring setup. So, it is feasible to explain the
two-particle quantum correlations according to the common cause principle in
a contextual manner.\bigskip

\textbf{Appendix\bigskip }

Let us assume that in three different stages of an experiment- hereafter
called first, second and third stages- Sam sends three pairs of colored
balls ($a,b$), ($a,c$) and ($c,b$) to both Alice and Bob. Then, after many
trials, Alice and Bob upload their registered results in distinct categories
(indicating the time and duration of the corresponding experiment) to a
remote computer which can perform statistical calculations. Each stage of
experiment is controlled by two complementary executive algorithms which are
chosen by Sam with equal chance. Hence, there are three equiprobable couples
of such executive algorithms: ($\mathcal{A}_{1}$, $\mathcal{A}_{2}$), ($%
\mathcal{A}_{1}^{\prime }$, $\mathcal{A}_{2}^{\prime }$), and ($\mathcal{A}%
_{1}^{\prime \prime }$, $\mathcal{A}_{2}^{\prime \prime }$). Either of the
two first algorithms $\mathcal{A}_{1}$ or $\mathcal{A}_{2}$ is used when Sam
decides to send $(a_{A}+,b_{A}\pm )$ joined with $(b_{B}\mp ,a_{B}-)$ balls
or $(a_{A}-,b_{A}\pm )$ joined with $(b_{B}\mp ,a_{B}+)$ balls to Alice and
Bob respectively, as was described in section 5. The second and the third
pairs of algorithms are characterized by replacing $b$ or $a$ colored balls
with $c$s in the above settings, correspondingly. Table \textbf{A}$_{\mathbf{%
1}}$ shows the probabilistic features of two algorithms $\mathcal{A}%
_{1}^{\prime }$ and $\mathcal{A}_{1}^{\prime \prime }$. It is assumed that
for both situations, when Sam executes any of the two algorithms $\mathcal{A}%
_{1}^{\prime }$ or $\mathcal{A}_{1}^{\prime \prime }$, the probability that
Bob records a positive result is equal to 0.04. For simplicity, we have not
mentioned the executive routines of $\mathcal{A}_{2}^{\prime }$ and $%
\mathcal{A}_{2}^{\prime \prime }$ which are simply the complements of two
earlier programs, respectively.\bigskip

\[
\begin{tabular}{|c|c|c|}
\hline
{\small Frequency}$\diagdown ${\small Results} & $%
(a_{A}+,c_{A}-;c_{B}+,a_{B}-)$ & $(a_{A}+,c_{A}+;c_{B}-,a_{B}-)$ \\ \hline
{\small Frequency according to }$\mathcal{A}_{1}^{\prime }$ & 0.04 & 0.96 \\ 
\hline
{\small Frequency}$\diagdown ${\small Results} & $%
(c_{A}+,b_{A}-;b_{B}+,c_{B}-)$ & $(c_{A}+,b_{A}+;b_{B}-,c_{B}-)$ \\ \hline
{\small Frequency according to }$\mathcal{A}_{1}^{\prime \prime }$ & 0.04 & 
0.96 \\ \hline
\end{tabular}
\]

\begin{center}
{\small Table }\textbf{A}$_{\mathbf{1}}${\small . The executive routines of }%
$\mathcal{A}_{1}^{\prime }${\small \ and }$\mathcal{A}_{1}^{\prime \prime }$ 
{\small in terms of some given probabilities.\bigskip }
\end{center}

When Sam executes, say, $\mathcal{A}_{1}^{\prime }$ algorithm, it may happen
that Bob decides to collect $b$ colored balls instead of $c$s in a given
course of experiment. For such a case, he does not record any result. In
other words, $\Pr (a_{A}+,b_{A}\pm ;b_{B}\mp ,a_{B}-|\ \mathcal{A}%
_{1}^{\prime })=0$. A similar situation may happen for Alice. So, in some
given courses, it may occur that Alice and/or Bob, because of their
unsuitable choices, register no result. Nevertheless, the experiment is
repeated for many times and one can assume that there are always situations
where Alice and Bob register simultaneously all transmitted colored balls in
a given course. Moreover, because in all executive algorithms, the
probability of occurring an event is supposed to be equivalent to the
frequency of occurrence of the same event (see footnote in section 5), a
remote computer can neglect the stages in which any of the two observers
detects no result and calculate the frequency of occurrence of the joint
outcomes in those stages where all joint data are registered. This can be
done by comparing the total sum of color-registered data with all that the
color-recognizer device detects as passing balls. The experiment is declared
to be terminated when Alice and Bob both have full registered data, i.e.,
when all color-registered balls match the passing balls for all triple
stages (see assumption 5 in section 5).\bigskip 

\[
\begin{tabular}{|c|cccc|}
\hline
{\small Statistical Calculations}$\diagdown ${\small Results} & $%
(a_{A}+;c_{B}+)$ & \multicolumn{1}{|c}{$(a_{A}+;c_{B}-)$} & 
\multicolumn{1}{|c}{$(a_{A}-;c_{B}+)$} & \multicolumn{1}{|c|}{$%
(a_{A}-;c_{B}-)$} \\ \hline
{\small Frequency for the second stage} & $\frac{1}{2}\times 0.04$ & 
\multicolumn{1}{|c}{$\frac{1}{2}\times $ $0.96$} & \multicolumn{1}{|c}{$%
\frac{1}{2}\times 0.96$} & \multicolumn{1}{|c|}{$\frac{1}{2}\times 0.04$} \\ 
\hline
{\small Statistical Calculations}$\diagdown ${\small Results} & $%
(c_{A}+;b_{B}+)$ & $(c_{A}+;b_{B}-)$ & $(c_{A}-;b_{B}+)$ & $(c_{A}-;b_{B}-)$
\\ \hline
{\small Frequency for the third stage} & $\frac{1}{2}\times 0.04$ & $\frac{1%
}{2}\times $ $0.96$ & $\frac{1}{2}\times 0.96$ & $\frac{1}{2}\times 0.04$ \\ 
\hline
\end{tabular}
\]

\begin{center}
{\small Table }\textbf{A}$_{\mathbf{2}}${\small . The remote computer
calculates the frequency of occurrence of a joint result for the second and
third stages.\bigskip }
\end{center}

For the second and third stages of this trilogy, the statistical
calculations of the remote computer are summarized in table \textbf{A}$_{%
\mathbf{2}}$. Here, all the complementary algorithm pairs ($\mathcal{A}_{1}$%
, $\mathcal{A}_{2}$), ($\mathcal{A}_{1}^{\prime }$, $\mathcal{A}_{2}^{\prime
}$) and ($\mathcal{A}_{1}^{\prime \prime }$, $\mathcal{A}_{2}^{\prime \prime
}$) constitute the field of common cause. This means that the corresponding
correlations disappear when any of the two algorithms in each pair is held
fixed. On the other hand, each of the above pairs is defined in a specific
context which is characterized by a definite pair of colored balls sent to
the observers. Disregarding the role of conditional contexts in the
application of the probability rules, one can derive some Bell-type
inequalities, such as (see, e.g., reference [13])

\begin{equation}
\Pr (a_{A}+;b_{B}+)\leq \Pr (a_{A}+;c_{B}+)+\Pr (c_{A}+;b_{B}+)  \tag{A-1}
\end{equation}

Putting the corresponding results from the tables 2 and \textbf{A}$_{\mathbf{%
2}}$ into the above inequality, we shall observe a clear inconsistency:

\begin{equation}
\frac{1}{2}\times 0.15\leq \frac{1}{2}\times 0.04+\frac{1}{2}\times 0.04%
\text{ Or }0.15\leq 0.08  \tag{A-2}
\end{equation}

The inconsistency, however, is removed, when one remembers that the
probability functions in relation (A-1) are derived under different
contexts. One such derivation is shown in the relation (25). Indeed, if we
apply the probability rules correctly, the Bell-type inequality (A-1) will
be never obtained.

\end{document}